\documentclass{ws-procs975x65}

\begin{document}
\title{Transport through Double-Dots coupled to normal and 
superconducting leads}

\author{Y. Tanaka and N. Kawakami}

\address{Department of Applied Physics, Osaka University,\\
Suita, Osaka 565-0871, Japan\\
E-mail: tanaka@tp.ap.eng.osaka-u.ac.jp}

\maketitle

\abstracts{
We study transport through double quantum dots coupled to normal and 
superconducting leads, where the Andreev reflection plays a key role in
determining characteristic transport properties.
We shall discuss two typical cases, i.e. double dots with serial or parallel geometry. For the parallel geometry, the interference of electrons 
via multiple paths is induced, so that the transmission probability 
has Fano-type dip structures which are symmetric with respect 
to the Fermi energy.
We also investigate the Aharonov-Bohm(AB) effect for the parallel geometry. 
In some particular situations, we find that the general AB period for 
double dots, 4$\pi$, is reduced to 2$\pi$.
}

\section{Introduction}
Recent advances in nanotechnology have enabled us to realize mesoscopic normal-metal/superconductor hybrid systems. In these systems, the Andreev reflection plays an important role for quantum transport. In particular, the Andreev reflection for a quantum dot coupled to normal and superconducting leads gives rise to characteristic transport properties due to the discreteness of energy levels in a dot.\cite{Beenakker,Zhao1,Sun1,Loss,Zhao2,Chen}
Moreover, the interplay between the Andreev reflection and the Kondo effect in quantum dot systems has been investigated 
intensively.\cite{Fazio,Schwab,Clerk,Cuevas,Sun2,Avishai,Graber}

In this work\cite{Tanaka}, we study transport through double quantum dots (DQD's) coupled to normal and superconducting leads. Here, we concentrate on transport due to the Andreev reflection, which is referred to the Andreev tunneling in the
following. We shall discuss two typical cases: the DQD is connected in series or parallel.
In particular, we focus on the interference effect, which is caused via multiple paths of electron propagation, on the Andreev reflection in 
the parallel DQD.\cite{Tanaka,Yong,Peng}
Since the interference effect is sensitive to the magnetic flux, we also investigate the influence of the Aharonov-Bohm(AB) effect on the Andreev tunneling.

In the following, we first give a brief explanation of the model, and then describe the results for the differential conductance due to the Andreev tunneling in Sec. \ref{sec:result}. A brief summary is given in the last section.

\section{Model}
\begin{figure}[h]
\begin{center}
\includegraphics[scale=0.4]{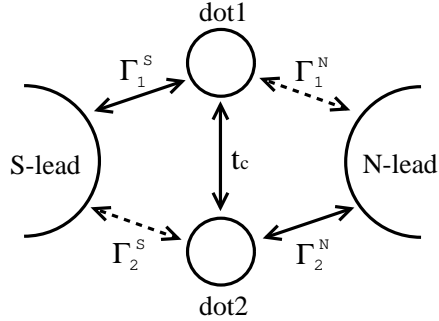}
\end{center}
\caption{
DQD system coupled to normal(N) and superconducting(S) leads. 
$t_c$ is the inter-dot tunneling, and 
$\Gamma^{N(S)}_{i} (i=1,2)$
represents the resonance width due to the transfer between 
the dot-$i$ and normal(superconducting) lead.
Note that two dots are connected in series
when $\Gamma^{N}_{1}= \Gamma^{S}_{2}=0$ (the dashed arrows),
which corresponds to $\alpha =0$ (see text).
}
\label{modelNS}
\end{figure}

We consider a DQD system coupled to normal and
 superconducting leads (N/DQD/S) shown in Fig. \ref{modelNS}. 
In this figure, two dots are coupled via the inter-dot 
tunneling  $t_c$, and the dot-$i$ is connected to the normal
(superconducting) lead via the tunneling, which causes the 
 resonance width of $\Gamma^{N(S)}_{i} (i=1,2)$.
Introducing the ratio $\alpha = \Gamma^{N}_{1}/\Gamma^{N}_{2}$
$=\Gamma^{S}_{2}/\Gamma^{S}_{1}$ in the same notation as in Refs. 
\refcite{Gueva}-\refcite{tanaka:2004},
we discuss two typical cases  $\alpha =0$ and $\alpha \sim 1$, namely,
  two dots are connected  in series or parallel.
We assume that the superconducting lead is well described by the BCS 
theory with a superconducting gap $\Delta$.
In addition, the intra-dot Coulomb interaction is ignored for simplicity.
Since we are interested in the Andreev tunneling, we concentrate on
the zero temperature case ($T=0$) in the region of
small bias voltage $V$ (i.e. $|V|<\Delta$).
We calculate the differential conductance $dI/dV$ 
($I$: current) and the density of states (DOS)
of the dots by using the Keldysh Green functions.

\section{Numerical Results} \label{sec:result}

\subsection{Andreev tunneling in serial and parallel DQD systems}
We first discuss the Andreev tunneling in the serial
and parallel DQD systems.
For simplicity, we fix the energy level of the dot-$i$ 
($\varepsilon_{i}$)
at the Fermi energy ($\varepsilon_{1}=\varepsilon_{2}=0$), 
and use the gap $\Delta$ as the unit of energy.

\begin{figure}[h]
\includegraphics[scale=0.4]{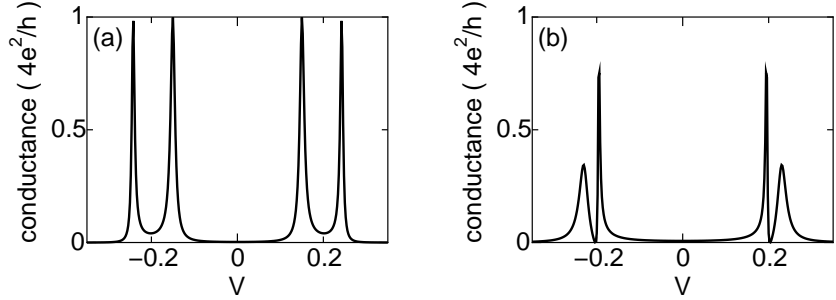}
\caption{
The conductance as a function
of the bias voltage $V$ for the N/DQD/S system.
(a) serial DQD ($\alpha=0$), (b) parallel DQD ($\alpha=0.7$).
We set $t_c=0.2, \Gamma^N_2=0.01, \Gamma^S_1=0.1$ ($\Delta$=1).
}
\label{conDQD}
\end{figure}
Figure \ref{conDQD} shows the differential conductance as a function
of the bias voltage $V$ for 
(a) serial ($\alpha=0$), (b) parallel ($\alpha=0.7$) DQD system. 
In both DQD systems,
the conductance has four peaks in its voltage dependence, 
which are symmetric with
respect to the Fermi energy. We shall discuss characteristic aspects of 
these peaks in terms of the DOS of the dots.
Note first that the interdot coupling $t_c$ forms the bonding 
and anti-bonding states
for electrons in dots, whereas those for holes are obtained by inverting 
the DOS profile with respect 
to the Fermi energy. The Andreev reflection at the 
DQD/S-lead interface mixes these states for electrons
and holes, giving rise to the four Andreev bound states in the dots.
Therefore, as mentioned above, the DOS of the dots has the four peaks,
where the width of these peaks is determined by the resonance width 
$\Gamma^{N}_{i}$.
In the serial DQD system, these peaks are indeed observed in the
profile of the conductance as shown in Fig. \ref{conDQD}(a). 

\begin{figure}[h]
\begin{minipage}{.45\linewidth}
\begin{center}
\includegraphics[scale=0.4]{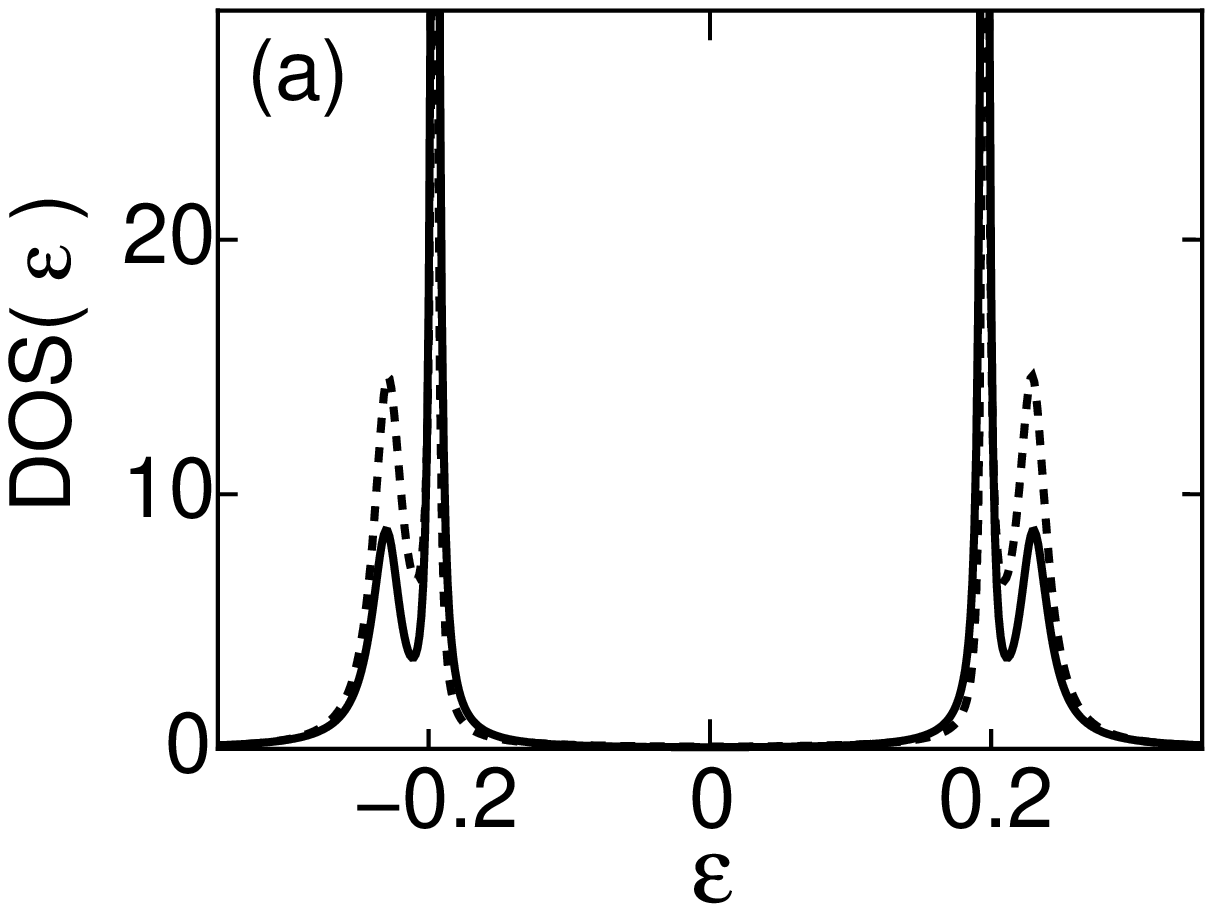}
\end{center}
\end{minipage}
\begin{minipage}{.45\linewidth}
\begin{center}
\includegraphics[scale=0.3]{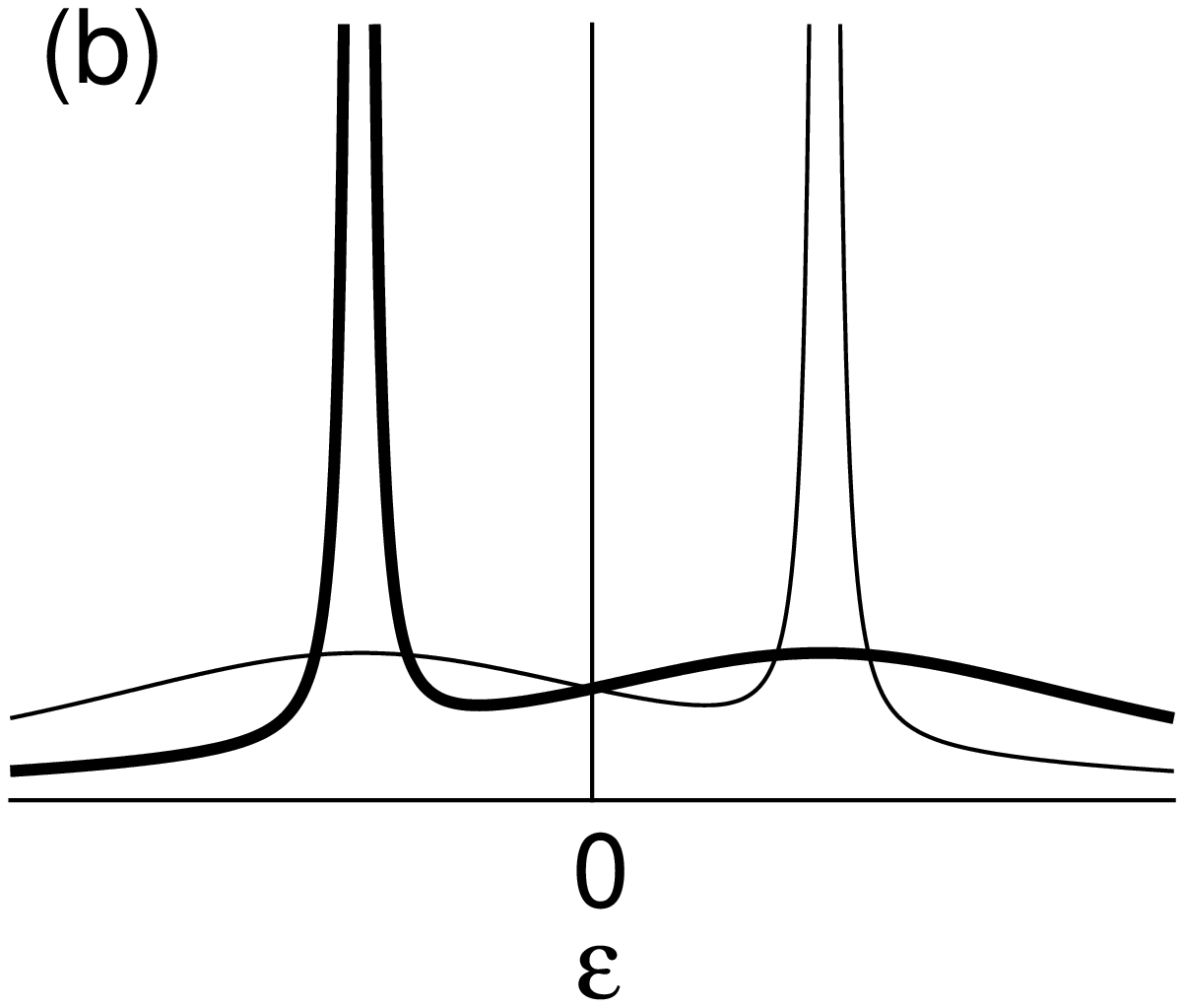}
\end{center}
\end{minipage}
\caption{(a) DOS of the dot for the N/parallel-DQD/S system. 
The solid (dashed) line is for the dot-1(2).
The parameters are the same as in Fig. \ref{conDQD}(b).
(b) Sketch of DOS for the  N/parallel-DQD/N system.
The thick (thin) line is for electrons (holes).
$\varepsilon=0$ corresponds to the Fermi energy.
}
\label{dos}
\end{figure}

On the other hand, in the parallel DQD,
Fano-type dip structures ($V \simeq\pm 0.2$) appear 
in addition to the four peaks, as shown in Fig. \ref{conDQD}(b).
Here, it is instructive to recall the interference effect
in the parallel DQD system coupled to two normal leads
(N/parallel-DQD/N). In this system,  the 
DOS for electrons in the dots consists of sharp and broad 
resonance peaks,
as shown by the thick line in Fig. \ref{dos}(b). 
When an electron transports via these resonances, 
its transmission probability acquires a Fano-type dip structure 
around the position where the sharp peak in the DOS is
 located.\cite{Gueva,tanaka:2005,tanaka:2004}
We note that the DOS for holes 
 is obtained by inverting that for electrons with respect 
to the Fermi energy (thin line in Fig. \ref{dos}(b)).
Coming back to the N/parallel-DQD/S system, we now see that
mixing of the electron and hole states induced by the
 Andreev reflection gives rise to 
four Andreev bound states in the dots, similarly to
the serial case. In contrast to the latter case, however, 
the resonances in the parallel case have two different widths: 
two sharp peaks and two broad peaks shown in Fig. \ref{dos}(a). 
The interference between the distinct transport channels via these resonances
gives rise to the Fano-type dip structures in the conductance, 
which are symmetric with respect to the Fermi energy
(Fig. \ref{conDQD}(b)).

Here, we make a brief comment on the special case $\alpha =1$, i.e. the symmetric couplings with leads. In this case, the sharp peaks in Fig. \ref{dos}(a)
 become of delta-function type, which means that the corresponding 
local states in the dots are completely decoupled from the leads. 
Therefore, the remaining states with the broad resonances in Fig. \ref{dos}(a)
only contribute to the Andreev reflection, so that the transport
 shows similar behavior to the case of 
a single dot coupled to normal and superconducting leads.

\subsection{AB effect in N/DQD/S system}

We next discuss how the Andreev tunneling changes its character
when the magnetic flux is added in the parallel DQD system.
Here, we assume that the magnetic flux equally pierces the two subrings 
formed by the interdot tunneling $t_c$, so that
the effect of the magnetic flux is symmetrically incorporated in
the tunneling between the dot and the lead.

Before considering the N/DQD/S system, we briefly mention
the AB effect in the N/DQD/N system.
As noted in the recent 
literature,\cite{Jiang:2002,Orellana:2004,Kang:2004,Zhi:2004}
the interdot coupling between the dots forms two-subring structure,
so that the AB period in the N/DQD/N system becomes $4\pi$
instead of the normal period of $2\pi$.
Similarly, the AB period in the N/DQD/S system is
expected to be $4\pi$ in general,
as pointed out by Zhang \textit{et al.}.\cite{Yong}
We find, however, that the AB period is reduced to $2\pi$
in some particular situations, as will be explicitly shown
below.

\begin{figure}[h]
\includegraphics[scale=0.4]{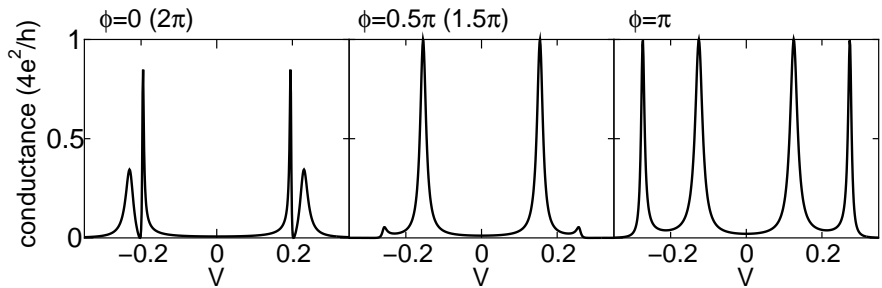}
\caption{
The conductance as a function of $V$ for several choices 
of the magnetic flux $\phi$.
The other parameters are the same as in Fig. \ref{conDQD}(b).
}
\label{conAB}
\end{figure}

Figure \ref{conAB} shows the conductance for several values of $\phi$, 
where $\phi$ represents the normalized value of the magnetic flux $\Phi$ as
 $\phi =2\pi\Phi/\Phi_0$ ($\Phi_0=h/e$).
The other parameters are the same as in Fig. \ref{conDQD}(b).
It should be noted that the conductance is symmetric with respect to 
the Fermi energy for
any value of $\phi$ and has the AB period of $2\pi$ (not $4\pi$).
More precisely, the conductance at the magnetic flux $\phi$, $G(\phi)$, 
 satisfies $G(\phi)=G(2\pi-\phi)$.
Including the situation for Fig. \ref{conAB}, we obtain
the general condition that reduces the AB period to $2\pi$,
\renewcommand{\labelenumi}{(\roman{enumi})}
\begin{enumerate}
  \item $\varepsilon_{1}=\varepsilon_{2}=0,\,
         (0<\alpha\le 1)$
  \item $\varepsilon_{1}=-\varepsilon_{2}(\neq 0)$
        and $\alpha=1$.
\end{enumerate}

The condition for the energy levels of the dots,
 $\varepsilon_{1}=-\varepsilon_{2}$,
means that the bonding state and the anti-bonding state in dots
are symmetric with respect to the Fermi energy
($\varepsilon= \pm \sqrt{\varepsilon_{1(2)}^{\,2}+t_c^2}\, $).
Then, the DOS of the dot-$1(2)$ at the magnetic flux $\phi$, 
$\rho_{1(2)}(\phi)$ has the AB period of $2\pi$
($\rho_{1(2)}(\phi)=\rho_{1(2)}(\phi+2\pi)$)
for the case (i).
On the other hand,
$\rho_{1(2)}(\phi)$ for the case (ii) has the periodicity of
$\rho_{1(2)}(\phi)=\rho_{2(1)}(\phi+2\pi)$,
namely, the DOS of the dot-$1(2)$ in the absence of the magnetic
flux is the same as that of the dot-$2(1)$ for $\phi=2\pi$.
Since the DOS with this periodicity directly determines electron
transport,
the conductance changes in the AB period of $2\pi$.

In this connection, we stress the difference in the AB period 
between our result and the related work 
by Zhang \textit{et al.}\cite{Yong},
who treated the same DQD system. We have found here that 
the AB period is reduced to $2\pi$ in the case (ii),
although they claimed that it still remains $4\pi$ in
 the same condition.

\section{Summary}

We have studied transport properties through the DQD
 coupled to normal and superconducting leads. It has been 
 discussed that the four-peak structure in the DOS is formed  
 by the bonding and antibonding states in the dots 
coupled to the superconducting lead. This structure 
in the DOS indeed determines the characteristic 
properties in the differential conductance, in accordance with
the results of  Zhang \textit{et al.}.\cite{Yong}
In particular, in the parallel DQD system, the interference effect 
due to the multiple paths
gives rise to the Fano-type dip structures in the differential conductance,
which are symmetric with respect to the Fermi energy.
We have also found the interesting fact  that 
the AB period is reduced to $2\pi$ in some particular situations, 
which is contrasted to the AB period  $4\pi$  expected generally
for DQD systems.


\begin{thebibliography}{99}

\bibitem{Beenakker} 
C. W. J. Beenakker,
Phys. Rev. B \textbf{46}, 12841 (1992).

\bibitem{Zhao1} 
H. -K. Zhao and G. v. Gehlen,
Phys. Rev. B \textbf{58}, 13660 (1998).


\bibitem{Sun1} 
Q. -f. Sun, J. Wang, and T. -h. Lin,
Phys. Rev. B \textbf{59}, 3831 (1999);
Phys. Rev. B \textbf{59}, 13126 (1999);
Phys. Rev. B \textbf{62}, 648 (2000).

\bibitem{Loss} 
P. Recher, E. V. Sukhorukov, and D. Loss,
Phys. Rev. B \textbf{63}, 165314 (2001).

\bibitem{Zhao2} 
H. -K. Zhao and J. Wang,
Phys. Rev. B \textbf{64}, 094505 (2001).

\bibitem{Chen} 
Z. Chen, J. Wang, B. Wang, and D. Y. Xing,
Phys. Lett. A \textbf{334}, 436 (2005).

%
%
\bibitem{Fazio} 
R. Fazio and R. Raimondi,
Phys. Rev. Lett. \textbf{80}, 2913 (1998).

\bibitem{Schwab} 
P. Schwab and R. Raimondi,
Phys. Rev. B \textbf{59}, 1637 (1999).

\bibitem{Clerk} 
A. A. Clerk, V. Ambegaokar, and S. Hershfield,
Phys. Rev. B \textbf{61}, 3555 (2000).

\bibitem{Cuevas} 
J. C. Cuevas, A. L. Yeyati, and A. Martin-Rodero,
Phys. Rev. B \textbf{63}, 094515 (2001).

\bibitem{Sun2} 
Q. -f. Sun, H. Guo, and T. -h. Lin,
Phys. Rev. Lett. \textbf{87}, 176601 (2001).

\bibitem{Avishai} 
A. Golub and Y. Avishai,
Phys. Rev. B \textbf{69}, 165325 (2004).

\bibitem{Graber}
M. R. Gr\"{a}ber, T. Nussbaumer, W. Belzig, and 
C. Sch\"{o}nenberger,
Nanotechnology \textbf{15}, S479 (2004).

%
%
\bibitem{Tanaka} Our work was also presented at the JPSJ meeting 
(September 19, 2005, Kyoto, Japan),
where we discussed the Andreev tunneling in both of serial and parallel
DQD systems. We note that
the parallel DQD case was already discussed by Zhang \textit{et al.}
 in Ref. \refcite{Yong}.
Our results are in accordance with theirs in this case.

\bibitem{Yong} 
Y. -P. Zhang, H. Yu, Y. -F. Gao, and J. -Q. Liang,
Phys. Rev. B \textbf{72}, 205310 (2005).

\bibitem{Peng} 
J. Peng, B. Wang, and D. Y. Xing,
Phys. Rev. B \textbf{71}, 214523 (2005).

\bibitem{Gueva} 
M. L. L. de Guevara, F. Claro, and P. A. Orellana,
Phys. Rev. B \textbf{67}, 195335 (2003).

\bibitem{tanaka:2005} 
Y. Tanaka and N. Kawakami,
Phys. Rev. B \textbf{72}, 085304 (2005).

\bibitem{tanaka:2004} 
Y. Tanaka and N. Kawakami,
``REALIZING CONTROLLABLE QUANTUM STATES'',
Proc. of the International Symposium on Mesoscopic Superconductivity and Spintronics 2004, pp.433-438 (World Scientific, Singapore, 2005).


\bibitem{Jiang:2002} 
Z. T. Jiang, J. Q. You, S. B. Bian, and H. Z. Zheng, 
Phys. Rev. B \textbf{66}, 205306 (2002).

\bibitem{Orellana:2004} 
P. A. Orellana, M. L. Ladr$\acute{\textrm{o}}$n de Guevara, and F. Claro, 
Phys. Rev. B \textbf{70}, 233315 (2004).

\bibitem{Kang:2004} 
K. Kang and S. Y. Cho, 
J. Phys.: Condens. Matter \textbf{16}, 117 (2004).

\bibitem{Zhi:2004} 
Z. -M. Bai, M. -F. Yang, and Y. -C. Chen, 
J. Phys.: Condens. Matter \textbf{16}, 2053 (2004).


\end{thebibliography}
\end{document}